\documentclass[prd,twocolumn,amssymb,amsmath,nofootinbib]{revtex4-1}

\usepackage{amssymb,amsmath,mathrsfs,epstopdf,slashed,color}
\usepackage{dcolumn}   
\usepackage{bm} 
\usepackage{tikz}
\usetikzlibrary{matrix}
\usetikzlibrary{positioning}
\usepackage{ifpdf}
\ifpdf
\usepackage{graphicx}
\usepackage{hyperref}
\else
\usepackage[dvipdfmx]{graphicx}
\usepackage[dvipdfmx]{hyperref}
\usepackage[numbers]{natbib}
\usepackage{hypernat}
\fi

\DeclareMathOperator*{\Tr}{Tr}

\begin{document}

\title{Baryon Number Violating Scatterings in Laboratories}

\author{S.-H. Henry Tye$^{1,2}$}
\email{iastye@ust.hk}
\author{Sam S.C. Wong$^{1,3}$}
\email{scswong@ust.hk}

\affiliation{$^1$ Department of Physics and Jockey Club Institute for Advanced Study, \\Hong Kong University of Science and Technology, Hong Kong}
\affiliation{$^2$ Department of Physics, Cornell University, Ithaca, NY 14853, USA}
\affiliation{$^3$ Department of Physics, Center for Theoretical Physics, Columbia University, New York, NY 10027, USA}

\date{\today}
  
\begin{abstract}
Earlier estimates have argued that the baryon number violating scattering cross-section in the laboratory is exponentially small so it will never be observed, even for incoming 2-particle energy well above the sphaleron energy of 9 TeV. 
However, we argue in Ref.\cite{tye:2015tva} that, due to the periodic nature of the sphaleron potential, the event rate for energies above the sphaleron energy may be high enough to be observed in the near future. That is, there is a discrepancy of about 70 orders of magnitude between the two estimates. Here we argue why and how the multi-sphaleron processes are crucial to the event rate estimate, a very important ``resonant tunneling" property that has not been taken into account before. We also summarize the input assumptions and reasoning adopted in our estimate, when compared to the earlier estimates.

\end{abstract}

\maketitle

\section{Introduction}

There is a large discrepancy in the predictions on the baryon plus lepton number ($B+L$) violating scattering cross-sections in the laboratory \cite{Hooft1976,Hooft1976a,Ringwald1990a,Espinosa1990,Porrati1990,
Khlebnikov1991b,Khoze1991,Khoze1991a,Mueller1991,Zakharov:1990dj,Gibbs1995a,Rebbi:1996zx,Rubakov1996,Bezrukov2003,Bezrukov2003a,Ringwald2003,Ringwald2003b,tye:2015tva}. We review where this huge discrepancy comes from and the input assumptions involved. We emphasize that a detailed quantum field theory (QFT) study is needed to fully resolve this discrepancy.

The electroweak theory is well established by now. With the $SU(2)$ gauge coupling $g\simeq 0.645$, or $\alpha_W=g^2/4\pi \simeq 1/30$, the W-boson mass $m_W=gv/2\simeq 80$ GeV (where $v=246$ GeV is the Higgs vacuum expectation value) and the Higgs mass $m_H=125$ GeV all measured, the theory (without extending it further) has no free parameter, so all dynamics are in principle completely determined. 
However, its non-perturbative properties remain to be fully explored. 
One important property is the sphaleron potential barrier height $E_{sph}=9.0$ TeV ($E_{sph}$ is also known as the sphaleron mass/energy), which separates vacua with different values of the Chern-Simons number $n=\mu/\pi$ \cite{Manton1983,Klinkhamer1984}. 

It is well known that the baryon number $B$ and the lepton number $L$ are not conserved in the electroweak theory \cite{Hooft1976,Hooft1976a}. So one likes to search for these ($B+L$)-violating processes in the laboratory, where $\Delta B=\Delta L= 3\Delta n$. Interesting parton (left-handed quarks) scatterings in proton-proton collisions are the $\Delta n \ne 0$ scatterings at quark-quark energy $E_{qq}>E_{sph}$; e.g., 
 a $\Delta n =-1$ quark-quark scattering goes like,
\begin{equation}\label{qqBL+1}
u_L + u_L \rightarrow e^+\mu^+ \tau^+ {\bar b}{\bar b}{\bar b}{\bar c}{\bar c}{\bar c}{\bar u} + X
\end{equation}
where $X$ includes particles that conserves $B$ and $L$ as well as the electric charge. So a single ($B+L$)-violating event can produce 3 positively charged leptons plus 3 ${\bar b}$-quarks. Other interesting possible experimental detections have also been proposed recently \cite{Ellis:2016ast,Brooijmans:2016lfv,Ellis:2016dgb}. 

 However, earlier estimates
have shown that such ($B+L$)-violating scattering cross-section $\sigma(E_{qq},{\Delta n} \ne 0)$ in the laboratory is exponentially small \cite{Espinosa1990,Ringwald1990a,Porrati1990,Khlebnikov1991b,Khoze1991,Mueller1991,Gibbs1995a,Rebbi:1996zx,Rubakov1996,Bezrukov2003,Bezrukov2003a,Ringwald2003,Ringwald2003b}; so that even if the quark-quark energy $E_{qq}$ is much higher than the sphaleron barrier height of 9 TeV, the event rate is still far too small to ever be observed in the laboratory \cite{Rubakov1996,Bezrukov2003a,Bezrukov2003,Ringwald2003,Ringwald2003b}.   Recently, we performed a different estimate and argue that the event rate may be high enough to be observed at the Large Hadron Collider (LHC) at CERN in the near future \cite{tye:2015tva}. Here we like to compare these two very different estimates. 

Let the cross section for usual electroweak (i.e., baryon number conserving) scattering at energy $E_{qq}$ be $\sigma_{EW}(E_{qq}, {\Delta n}=0)$. Let the total $q_Lq_L$ electroweak cross section at energy $E_{qq}$ be $\sigma_T (E_{qq}) = \sigma_{EW}(E_{qq},{\Delta n}=0)+\sigma(E_{qq},{\Delta n} \ne 0)$. We define the fraction of the ($B+L$)-violating processes among all electroweak processes to be 
\begin{equation}
\label{kappa}
\kappa (E_{qq})= \frac{\sigma(E_{qq},{\Delta n}\ne 0)}{\sigma_T(E_{qq})}\simeq \frac{\sigma(E_{qq},{\Delta n}=\pm 1)}{\sigma_{EW} (E_{qq}, {\Delta n}=0)} 
\end{equation}
where we ignore the other (${\Delta n} \ne 0, \pm 1$) contributions for simplification. For $E_{qq}$ below the sphaleron energy $E_{sph}=9.0$ TeV, $\kappa$ is exponentially small due to tunneling suppression. For any foreseeable $E_{qq}>E_{sph}$, it was argued earlier that $\kappa$ is still exponentially suppressed \cite{Espinosa1990,Porrati1990,Ringwald1990a,Khlebnikov1991b,Khoze1991,Mueller1991,Gibbs1995a,Rebbi:1996zx,Rubakov1996,Bezrukov2003,Bezrukov2003a,Ringwald2003,Ringwald2003b}, estimated to be of order 
\begin{equation}
\label{oldk}
\kappa (E_{qq}>E_{sph}) < 10^{-70}
\end{equation}
One may understand this phenomenon in a quantum mechanical (QM) model as due to energies being transferred to the ($B+L$)-conserving direction so the ($B+L$)-violating direction is left with $E < E_{sph}$, and one still has to go through barrier tunneling. As reviewed below, this energy transfer takes care of the ``few-to-many" suppression. 

We agree with this estimate for a single sphaleron. However, we claim that multi-sphaleron processes can drastically change the picture. The effective QM ($B+L$)-conserving degrees of freedom impact only the pre-factor of the tunneling rate in our QM model. 
When $E > E_{sph}$, i.e., the exponential tunneling suppression factor vanishes, the ($B+L$)-violating degree of freedom is described by a simple plane wave, and the pre-factor yields only a phase space factor. For the 14 TeV proton-proton energy, in addition to the parton distribution suppression factor (which is around $10^{-6}$), the phase space suppression factor from the pre-factor yields
\begin{equation}
\label{estimatek}
\kappa (E_{qq}>E_{sph}) \simeq \frac{2^{d/2}}{d}\left(1- \sqrt{\frac{E_{sph}}{E_{qq}}}\right)^{d/2} \sim 10^{-3}
\end{equation}
which is quoted in Ref.\cite{tye:2015tva} for $d \sim 4$. 

So there is a huge discrepancy between this value and that in Eq.(\ref{oldk}). It is clear that both estimates involve assumptions based on intuitions as well as approximations remaining to be fully justified. Since our result (\ref{estimatek}) sounds counter intuitive with respect to the conventional wisdom, we shall explain our argument in some detail. The key point is the resonant tunneling (or simply the resonance) phenomenon, which is present only in multi-sphaleron processes. Pictorially, this  point is summarized in FIG \ref{WtoSph}. To quantify this picture, we reduce the electroweak theory to a QM system, where the Bloch wave solution for a periodic sphaleron potential captures this resonance phenomenon.
This paper provides the resoming/justification for our Bloch wave treatment and the estimate of $\kappa$ in Ref.\cite{tye:2015tva}. In particular, the input/assumptions going into the analysis is presented for further examinations. 

The rest of the paper goes as follows. Sec. \ref{review} contains a brief review of our QM analysis. Sec. \ref{fewtomany} reviews the ``few-to-many" argument which provides an intuitive picture for the earlier estimate. We agree that there is a ``few-to-many" suppression factor in the quark-quark scattering concerning the sphaleron. This  ``few-to-many" exponential suppression factor together with the tunneling suppression for the single sphaleron case is briefly reviewed in Sec. \ref{single}. We believe that this result is qualitatively correct for the single (not multiple) sphaleron case, that $\kappa$ (\ref{oldk}) is exponentially small according to existing opinion even for $E > E_{sph}$. Sec. \ref{multiple} presents our key motivation and argument. Together with an old argument repeated in the Appendix, we explain why multi-sphaleron processes are very important. First, the multi-sphaleron processes are not multiply suppressed. Instead, on average, they should only be singly suppressed (comparable to that for a single sphaleron), while under the right condition (the resonant tunneling or resonance phenomenon), the process may be faster than that for a single sphaleron. This is our main difference from the existing belief. 
Instead of working directly with the electroweak theory, we propose in Sec. \ref{qftqm}  a reduction of the electroweak theory to a QM model for this process, which is studied in Ref.\cite{tye:2015tva}.
Sec. \ref{Bloch} explains the direction of the periodic sphaleron potential for the QM setup. Because of the presence of fermions, this direction for the Chern-Simons variable is different from that for the $|\theta\rangle$ vacuum. Here we assume the fermions are massless. We then consider the Bloch waves for energies above the sphaleron energy in Sec. \ref{Above}. These Bloch waves are simply free plane waves in QM.
Sec. \ref{massfermion} introduces the masses for the fermions, where we argue that the overall feature we are interested in, namely the ($B+L$)-violating scattering above the sphaleron energy,  is only slightly modified. 
Sec. \ref{twodim} discusses the number of effective QM degrees of freedom when the QFT problem is reduced to a QM problem. 
Here we attempt to extract this information from the QM models studied in the literature. This crude order-of-magnitude analysis suggests that there are about $d \sim 4$ such effective ($B+L$)-conserving dimensions. 
Sec. \ref{kappa1} discusses our estimate of the value of $\kappa$ quoted in Eq.(\ref{estimatek}) and in Ref.\cite{tye:2015tva}. This is clearly different from the exponentially small $\kappa$ value in Eq.(\ref{oldk}). The discussion is presented in Sec. \ref{discussion} and Sec. \ref{summary} gives a summary of our picture versus the prevalent picture. The Appendix reviews an old argument why the resonant tunneling phenomenon, well understood in QM, should be present in QFT.

\section{Brief Review} \label{review}

Let the SU(2) gauge and Higgs fields take field configurations ${\bf A}_{\nu} ({\bf x}, \mu(t))$ and $\Phi ({\bf x}, \mu(t))$ with sphaleron solutions at ${\bf x}_i$ at time $t_i$, where $i=1,2,...$ Integrating out the sphaleron solutions yields the one-dimensional quantum mechanical (QM) system for $\mu(t)$ \cite{Manton1983,tye:2015tva},
 \begin{eqnarray}
 \label{system}
 L &=&  \frac{m_{\mu}}{2}\dot{\mu}^2 -V(\mu)  \nonumber\\
 m_{\mu} & = & {44~ {\rm TeV}}/{g^3 v^2} \\
 V(\mu) &=& {4 ~{\rm TeV}} \left(\sin^2 \mu + 0.46 \sin^4 \mu \right)/g \nonumber
 \end{eqnarray}
 where integer values of the dimensionless dynamical variable $\mu/\pi$ can be identified with the Chern-Simons number, $n=\mu/\pi$ \cite{Klinkhamer1984} or the Hopf invariant \cite{Tye:2016pxi}. Here, the gauge coupling $g$ is also explicitly displayed. This periodic potential has a height of 9.1 TeV before the hyper-charge coupling is turned on, which will lower it to about 9.0 TeV. Note that the baryon number is different at different integer values of $n$, indicating that the potential is not that in a circle but truly periodic \cite{Bachas:2016ffl}. 

Since the sphaleron potential is actually periodic with respect to $\mu(t)$, we are led to consider Bloch wave solutions.  
Now $\kappa=0$ within a band gap (no Bloch wave solution), while it is ``tunneling" unsuppressed within a pass band (Brillouin zone). As the band gaps (around 70 GeV) are much wider than the band widths at low energies, averaging over a range of energies (say a hundred GeV) that include a couple of bands will yield the well-known WKB tunneling suppression factor \cite{tye:2015tva}, in agreement with earlier estimates.
On the other hand, for $E_{qq} > E_{sph}$, the Bloch wave is, to a very good approximation, simply a single mode of plane wave; so $\kappa$ is not tunneling suppressed. In Ref.\cite{tye:2015tva}, we guesstimate the $\kappa$ value (\ref{estimatek}) for the LHC 14 TeV proton-proton energy run in the near future, in addition to the parton distribution suppression factor. 
Comparing this to the bound (\ref{oldk}), we see that there is a roughly 70 orders-of-magnitude discrepancy between the two estimates.

Here we like to explain our estimate of $\kappa$, which is merely a (power-like) phase space suppression factor. Admittedly, the estimate (\ref{estimatek}) is only an order of magnitude guesstimate; nevertheless, its value strongly suggests that ($B+L$)-violating events may be observed in the laboratory, either at the LHC run at 14 TeV, or in future high energy proton-proton colliders. 
This huge discrepancy between (\ref{oldk}) and (\ref{estimatek}) boils down to whether $\kappa(E_{qq}>E_{sph})$ contains an exponential suppression factor or not. 

\section{The ``Few-to-Many" Factor} \label{fewtomany}

Numerical study \cite{Bezrukov2003} and theoretical arguments indicate that the ($B+L$)-violating cross-section, or $\kappa$, goes like, 
\begin{equation}
\label{2tomany}
 \kappa \sim e^{-4\pi/\alpha_W} \rightarrow e^{-2\pi/\alpha_W}
  \end{equation}
as the incoming 2-particle scattering energy $E \simeq 0$ increases to $E \gg E_{sph}$, thus yielding the approximate bound (\ref{oldk}). The first exponential suppression factor is the well-known instanton tunneling suppression, while the second suppression factor is the so called ``few-to-many" factor.
 
As argued at times in the literature (see e.g., Ref.\cite{Mattis:1991bj} and recently in \cite{Funakubo:2016xgd}), a big part of the exponential ``few-to-many" suppression factor (\ref{oldk}) may be understood in the following way.  Intuitively, a sphaleron, like a soliton, can be created from a coherent set of many fields, say of order $A/\alpha_W$ of them, where $A$ is a positive constant of order of unity. Starting from a 2 W-boson scattering, creating each additional W-boson needs a power of gauge coupling $g$ in the scattering amplitude, or a power of $\alpha_W$ in the cross-section. To reach that many (virtual and/or real) W-bosons, we need to go to $\alpha_W^{A/\alpha_W}$ order perturbatively in the cross-section, resulting in a suppression factor of 
 \begin{equation}
 \label{efactor}
 \exp \left(\frac{-A|\ln \alpha_W|}{\alpha_W}\right)
 \end{equation}
 for $\alpha_W \ll 1$. 
Although this exponential suppression factor mimics a tunneling suppression, it is expected to be present even if there is enough incoming energy to overcome the tunneling barrier height; that is, this suppression factor is due to the conversion of point-particle energy to create a soliton-like object made up of a coherent set of gauge fields (and Higgs bosons). Applying this intuitive argument to the sphaleron, one may naively argue that the ($B+L$)-violating process remains to be exponentially suppressed even for incoming 2-body energies way above the sphaleron energy.
 
 \begin{figure}[h]
 \begin{center}
  \includegraphics[scale=0.44]{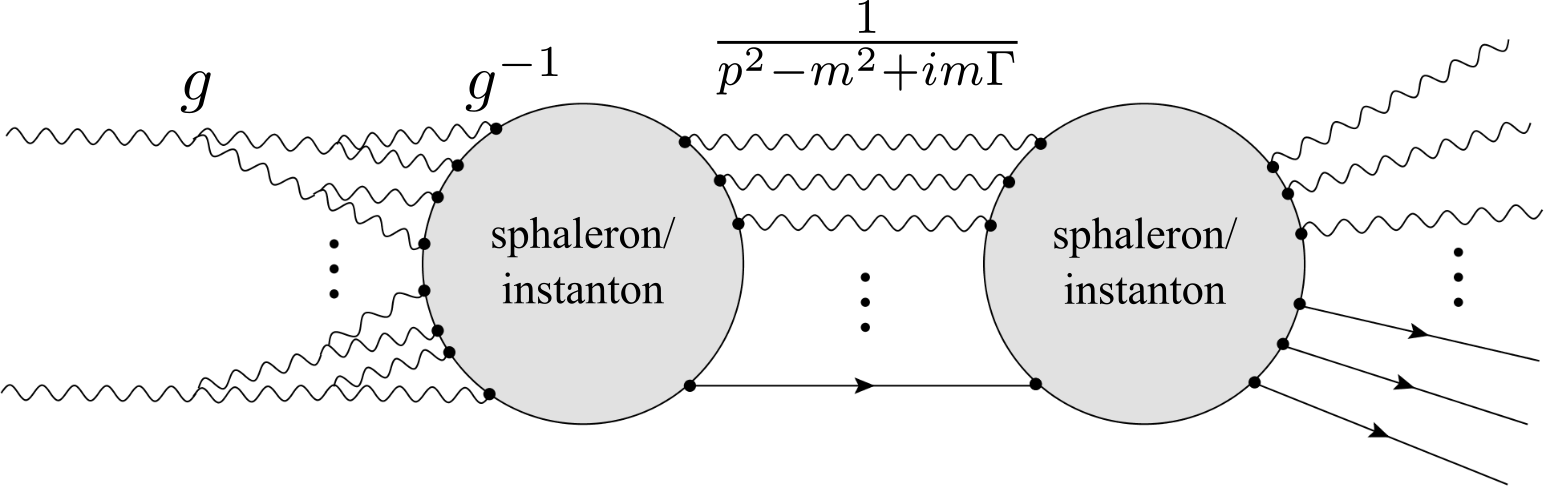}
  \caption{An example of how two hard incoming W-bosons  can give up their energy to a sphaleron. 
In the ``few-to-many" process, each additional W-boson costs a power of the gauge coupling $g$ while its coupling to the left sphaleron (denoted by the shaded disc) gains back a factor of $g^{-1}$.
 For a W-boson propagating between 2 sphalerons, we gain a factor of $1/g^2$ for an off-shell boson and a factor of $1/g^4$ for an on-shell boson.  The solid lines stand for the fermions.
  }
  \label{WtoSph}
 \end{center}
\end{figure}

However, the above argument is somewhat mis-leading. Let $A_{cl}(x)$ and $\Phi_{cl}(x)$ be the single classical sphaleron solution (all spin-isospin indices suppressed), where $A_{cl}(x) \propto e^{-m_Wr}$ and $\Phi_{cl}(x) \propto e^{-m_Hr}$ at large spatial distance $r$ from the sphaleron. Let $\delta A$ and $\delta \Phi$ be the fluctuations about this classical solution,
\begin{equation}
{\bf A}(x)= A_{cl}(x) + \delta A(x), \quad \quad
{\bf \Phi}(x) = \Phi_{cl}(x) + \delta \Phi(x)
\end{equation}
Naively, one inserts the fluctuating fields into the relevant Green's function for a scattering process involving $k$ fermions, $n_W$ W-bosons and $n_H$ Higgs fields,
\begin{equation}
\left\langle\psi(y_1) . . . \psi(y_{k})A(x_1) . . . A(x_{n_W}) \phi(z_1) . . .\phi(z_{n_H}) \right\rangle
\end{equation}
where $A(x_i)=\delta A(x_i)$ and $\phi(z_j)= \delta \Phi(z_j)$.
It was observed by Ringwald \cite{Ringwald1990a} and Espinosa \cite{Espinosa1990} that, to the leading order approximation, the Green function for a ($B+L$)-violating process involving $n_W$ W-bosons and $n_H$ Higgs bosons is given by
\begin{equation}
\left\langle\psi_1(y_1) . .  \psi_{12}(y_{12})A_{cl}(x_1) . .  A_{cl}(x_{n_W}) \Phi_{cl}(z_1) .  .\Phi_{cl}(z_{n_H})\right\rangle
\end{equation}
That is, it is the classical fields themselves that enter into the Green's function. 
Focusing on the gauge coupling $g$, we see that the gauge-fixed $A_{cl}(x) \propto 1/g$ in the constrained instanton solution \cite{Ringwald1990a},
\begin{eqnarray}
 &&A^{cl}_{\mu}(x) \\
 &=& \frac{i}{g}\left\{ \begin{array}{lc}
  -2 \frac{\rho^{2} \bar{\sigma}_{\mu\nu}  x_{\nu} }{x^2(x^2 + \rho^{2})} +{\cal O}((\rho v)^2) \quad  & |x_E|<m_W^{-1} \nonumber\\
  4\pi^2 \rho^2 \bar{\sigma}_{\mu\nu} \partial_{\mu}G_{m_W}(x) + {\cal O}(\rho^4 v^2) \quad  &|x_E| \ge m_W^{-1}
\end{array} \right.,  \nonumber 
\end{eqnarray}
where $G_{m}(x) = {m K_1(m|x|)}/(4\pi^2 |x|)$. For a sphaleron, we expect $\rho \sim 1/m_W$. This means that, for a W-boson to contribute to the formation of a sphaleron in the scattering amplitude, there is a power of $1/g$ from its coupling to the sphaleron, canceling the power of $g$ in its production, as illustrated in FIG {\ref{WtoSph}}. So converting 2 hard W-bosons to many soft W-bosons to form the sphaleron does not automatically lead to the exponential suppression factor (\ref{efactor}). Of course, this does not imply that, besides the tunneling suppression factor, there is no other suppression factors lurking around. Overall, we believe $A< 1$ is a reasonable estimate for a single sphaleron.
Note that, if nothing else happens, even if $A < 1$, the tunneling suppression factor (\ref{efactor}) (for example, $A\sim 0.3$) may still be enough to render the ($B+L$)-violating processes unobservable in the laboratory.

\section{Single Sphaleron Suppression}\label{single}

\begin{figure}[!h]
	\centering
	\includegraphics[width=0.45\textwidth]{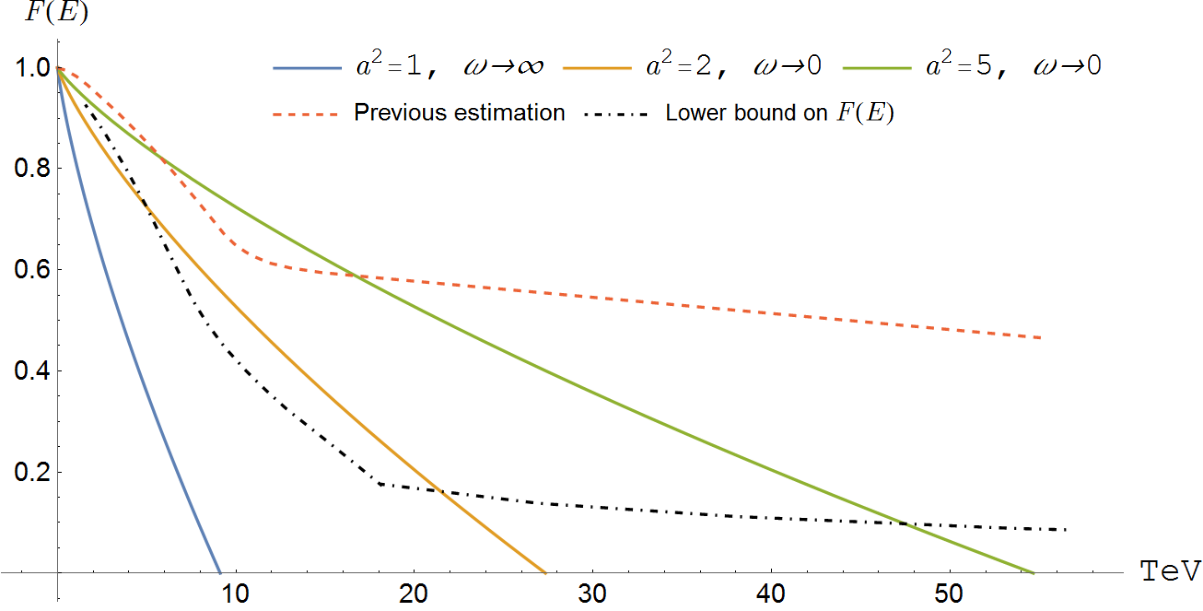}
	\caption{The magnitude of $F_0 = -\alpha_W \log (T)/2 \pi$, 
   where $T$ is the transmission coefficient, as a function of the simple harmonic oscillator frequency $\omega$ and the energy $E$. There is no tunneling suppression when $F_0=0$. The blue curve is for the $a=1, \omega \rightarrow \infty$ case, where $F_0 =0$ at $E=E_{sph}$, which roughly coincides with the $a=0, \omega \rightarrow 0$ case. The orange curve is for the $a^2=2, \omega \rightarrow 0$ case where $F_0 =0$ at $E=3E_{sph}$ \cite{Bonini:1999kj}.  The solid green curve is for the $a^2=5, \omega \rightarrow 0$ case where $F_0 =0$ at $E=6E_{sph}$. The red dash curve and black dot-dashed lower bound are from the numerical calculation in Ref.\cite{Bezrukov2003}.}
   \label{transmission}
\end{figure}

The above  ``few-to-many" suppression picture has been quantified to some extent. The energy dependence of ($B+L$)-violating cross-section at low energies $E \ll E_{sph}$ in the laboratory is estimated for a single sphaleron \cite{Espinosa1990,Ringwald1990a,Porrati1990,Khlebnikov1991b,Khoze1991,Mueller1991} to be 
\begin{eqnarray} \label{crosssec}
\sigma &\sim & \exp\left\{-\frac{4\pi}{\alpha_W}F(E)\right\}  \\
F(E) &\simeq & 1-\frac{9}{8}\bigg(\frac{E}{E_0}\bigg)^{4/3}+\frac{9}{16} \bigg(\frac{E}{E_0}\bigg)^{2} +{\cal O}\bigg(\bigg(\frac{E}{E_0}\bigg)^{8/3}\bigg) \nonumber
\end{eqnarray}
where $E_0=  \sqrt{6} \pi m_W/\alpha_W \simeq 15$ TeV.
In the work of Bezrukov et al. \cite{Bezrukov2003,Bezrukov2003a}, an estimate of the $(B+L)$-violating cross section at energy higher than the sphaleron barrier was performed. The basic idea behind is to solve the classical equations of motion for spherically symmetric $A_{\mu}$ and $\Phi$ along the following path in the complex plane of time,
\begin{equation}
 (t=-\infty+iT) \to (t=iT) \to (t=0) \to (t=+\infty).
 \end{equation}
The middle path from $(t=iT) \to (t=0) $ is Euclidean such that it corresponds to tunneling under the barrier. The field solution along the path determines the total incoming energy $E$ and particle number $N_i$, and also the suppression factor $\sigma\sim \exp[-\frac{4\pi}{\alpha_W}F(E,N_i)]$ coming from the Euclidean path. The numerical results shows that for small number of incoming particles $N_i\to 0$, the exponent function $F_0(E)=F(E,0)$ roughly reaches ${1}/{2}$ even at very high energy. Therefore, it was concluded that $(B+L)$-violating processes are highly suppressed even at $200$ TeV, as shown by the red dash curve in FIG \ref{transmission}. This yields the ``few-to-many" suppression factor (\ref{2tomany}), which leads to the bound (\ref{oldk}).

Another point of view by Ringwald \cite{Ringwald2003b} is to take the lower bound of $F_0(E)$ (i.e., upper bound of the cross-section) from the above work \cite{Bezrukov2003},  as shown by the black dot-dashed curve in FIG  \ref{transmission}, and combine it with the s-wave unitarity to put an upper bound on the ($B+L$)-violating cross section. Within this bound, there is still a possible region that the $(B+L)$-violating process may be observed at energies $E \gtrsim 70$ TeV in the next generation collider. Even with this very optimistic estimate, it is still below our estimate for the cross-section. We argue that the cross-section will reach (or get close) to the unitarity bound faster than their estimates.

\section{Multiple Sphaleron Processes}\label{multiple}

Naively, tunneling through two sphalerons is doubly suppressed with respect to tunneling through a single sphaleron. For this reason, multiple sphaleron processes have not been studied in any detail in the literature. First, we like to point out that tunneling through two sphalerons is only singly suppressed, not doubly suppressed (see the old argument in the Appendix). Next, we argue in the Appendix that the resonant tunneling phenomenon is responsible for this singly suppressed result, both in QM, which is well-known, and in QFT, which is much less appreciated. 

Let us consider FIG \ref{WtoSph}. For each W-boson leaving the first sphaleron and ending in the second sphaleron, we gain a factor of $1/g^{2}$, due to its coupling to the two sphalerons. This property suggests that multiple sphaleron processes are not as suppressed as one naively expects. Furthermore, for incoming energy $E \gg m_W$, there can be up to $E/m_W$ number of bosons going on-shell.
Each boson propagator reaching the resonance pole goes like
$$\frac{1}{p^2-m_W^2 + i m_W \Gamma_W} \rightarrow \frac{-i}{m_W\Gamma_W} \propto \frac{1}{\alpha_W}$$
since the W-boson decay width goes like $\Gamma_W \sim \alpha_W m_W$. So we gain a factor of $1/\alpha_W$ when the W-boson hits the resonance pole. That is, each on-shell W-boson between the two sphalerons can provide a factor of $1/\alpha_W^2$ to the process. We interpret this to be a resonant tunneling phenomenon when all (or a large enough number) of the intermediate particles hit their respective on-shell poles while $E < E_{sph}$. 

With $E \sim m_W/\alpha_W$, we can easily gain on average a factor (with $D$ a positive constant of order unity)
\begin{equation}
\label{Dfactor}
 \alpha_W^{-D/\alpha_W} \sim \exp \left(+\frac{D |\ln \alpha_W|}{\alpha_W} \right)
\end{equation}
In this simple  exercise of power counting of the coupling $\alpha_W$, $D$ in Eq.(\ref{Dfactor}) can be bigger than $A$ in Eq.(\ref{efactor}). 
This picture illustrates that double (and multiple) sphaleron processes not only are not doubly suppressed, but can, under the right circumstances, actually enhance the tunneling processes relative to the single sphaleron process. A few comments are in order here: \\
$\bullet$ Note that this ``resonance" effect will be absent if we perform the tunneling process by rotating to Euclidean time. Such a ``Euclidean time" calculation will naively yield a multiply suppressed tunneling rate, which suggests that multi-sphaleron processes are multiply suppressed and so totally negligible. \\
$\bullet$ The enhancement factor due to the resonance phenomenon should persist even for $E > E_{sph}$. In fact, higher energy allows more on-shell bosons, allowing further enhancement. Strictly speaking,  this resonance phenomenon is no longer tunneling related, but is intimately related to the existence of two (or more) barriers. \\
$\bullet$ The resonance phenomenon is present as long as there is an ``infinite" sum of  coherent paths to enhance the rate. Here, the $\Delta n = \pm 1$ ($B+L$)-violating process is enhanced by the presence of the second sphaleron without having to go through it. That is, the second sphaleron plays a catalytic role in enhancing the single sphaleron (i.e., $\Delta n = \pm 1$)  process. This possibility has been illustrated with the so called catalyzed tunneling example \cite{Tye:2009rb,Tye:2011xp} in a similar but different context.


\section{Reduction to a QM problem}\label{qftqm}

So it is important to consider multiple sphaleron processes in some detail. To study the multi-sphaleron processes in the electroweak theory is clearly very challenging. Since we are mostly interested in how the resonance phenomenon impact on the ($B+L$)-violating rate, we like to reduce the problem to a simpler problem without losing this phenomenon. In principle, we can cast the electroweak theory in the functional Schr\"odinger form, identify the ($B+L$)-violating direction as the most probable escape path (MPEP) \cite{coleman:1977py}, integrate out the fields and then drop the degrees of freedom (d.o.f.) that are not directly relevant to the exponential suppression factor. Since the ($B+L$)-conserving d.o.f. are orthogonal to the MPEP, they may contribute to the pre-factor in the tunneling rate, but not the exponent \cite{Tye:2009rb,Tye:2011xp}. The only d.o.f. we keep now is the Chern-Simons number for the gauge field, which is responsible for the ($B+L$)-violation, and we shall come back later to estimate the number of ($B+L$)-conserving d.o.f. and discuss their impact on the  ($B+L$)-violating rate. In contrast, as we shall review below, in some of the earlier versions of QM models, the ($B+L$)-conserving d.o.f. is not orthogonal to the ($B+L$)-violating direction. So the resulting physics can be quite different.

Instead of the Chern-Simons number, we could have chosen to keep the Hopf invariant for the Higgs field instead. Since the gauge and the Higgs fields couple closely to each other for the sphaleron solutions, we can identify the Chern-Simons number with the Hopf invariant \cite{Tye:2016pxi}. The end result yields a one-dimensional QM system with the Chern-Simons number as the dynamical quantum variable obeying the Schr\"odinger equation \cite{tye:2015tva}. As the multiple-sphaleron feature yields a periodic potential, the discrete translational symmetry allows us to solve it easily with the Bloch waves. Since the tunneling suppression is still rather severe for $E <E_{sph}$, we focus on $E \gtrsim E_{sph}$, and the Bloch wave solution allows us to go smoothly from  $E <E_{sph}$ to $E \gtrsim E_{sph}$. Our estimate of the ($B+L$)-violating scattering rate is based on this analysis. 

We can see the above picture in another way. A soliton has size $1/m_W$ while 2 hard particles colliding with enough energy to create it must have energy of order $m_W/\alpha_W$, so the time involved is correspondingly very short. However, as suggested in Ref.\cite{Li:1991xq}, the sphaleron case is different. Although the instanton, which is closely related to the sphaleron, is localized in Euclidean time, the corresponding Lorentzian time needs not be as localized. Although the spatial size of a sphaleron is small, the Lorentzian time needed to overcome the sphaleron barrier can be quite long. 
Let us express the sphaleron solution in terms of the spatial coordinates $\vec x$ and a parameter $\mu$, so $A_{cl}(x)=A_{cl}({\vec x}, \mu)$ and $\Phi_{cl}(x)=\Phi_{cl}({\vec x}, \mu)$ \cite{Manton1983}. Here we may treat $\mu$ as a function of time. For example, a typical single instanton solution treats $\mu\simeq \hat t$ for Euclidean time $\hat t$ with a $SO(4)$ symmetry. Here we shall stay in Lorentzian spacetime.

Recall the Chern-Simons number $n$. Going from $n=\mu/\pi=0$ (at $t = -\infty$) to $n=\mu/\pi=1$ (at $t \rightarrow +\infty$) implies that the universe goes from a vacuum with baryon number $B=0$ to another vacuum with baryon number $B=3$. To get an idea of the actual time involved in this transition, we like to lift the parameter $\mu(t)$ to be time-dependent.  
Treating $\mu(t)$ as a dynamical variable, one obtains the Lagrangian (\ref{system}).
One can then write down the corresponding one-dimensional quantum mechanical system and estimate the time it takes to go from $n=0$ to $n=1$. Once we have the one-dimensional QM setup, the analysis becomes straightforward. It is the interpretation of the result that takes some care. 

For energies below the barrier height,
Bloch wave solutions reflect the resonant tunneling phenomenon, which emerges when there is a coherent sum of infinite number of paths in the path integral formalism. (It is absent in the single sphaleron case.) Such a feature is difficult to capture numerically in QM, not to mention in QFT. However, we claim this is precisely where the discrepancy lies. Each field configuration for multiple sphalerons denotes  a most probable escape path in the field space. Integrating out the field space leaving $\mu(t)$ as the only remaining dynamical variable in a QM system allows us to see explicitly how this coherent sum of paths emerges. Clearly, it will be nice to study this phenomenon directly in QFT.  Resolving the $\kappa$ discrepancy puzzle is surely challenging. 
Here, we discuss this 70 orders of magnitude discrepancy in some detail in the context of QM and explain our estimate of $\kappa$ quoted in Ref.\cite{tye:2015tva}.

The earlier argument that it is exponentially small is in part based on an examination of an analogous two-dimensional  QM model which supposedly captures the key features of the actual QFT phenomenon \cite{Bonini:1999cn,Bonini:1999kj,Bezrukov:2003tg}. The QM model has in the baryon number violating direction a single potential barrier that mimics the sphaleron potential. The other direction mimics ordinary baryon number conserving channels. As enough energy of an incoming wave is diverted to the baryon number conserving direction, the tunneling suppression takes place in the baryon number violating direction even if one starts with an initial energy higher than the barrier height. The result is checked by a numerical study of the actual electroweak field theory case \cite{Bezrukov2003,Bezrukov2003a}. 
This analysis is based on a single sphaleron potential barrier and we do not disagree with the conclusion. 
The difference emerges in  the two (or higher) dimensional QM case, as we shall explain later. 

 
 \section{The Direction for the Chern-Simons Variable}\label{Bloch}
 
 Here, we shall provide some background and clarifying discussions to our Bloch wave analysis \cite{tye:2015tva}. 
 Although the discussion there is strictly for the case with no fermions, ($B+L$)-violation is included in the discussion on phenomenology. The presence of fermions (with their zero modes) changes the picture in a fundamental way.  Here we shall be more precise on this point.
 Ref.\cite{tye:2015tva} starts with the Schr\"odinger equation for the Chern-Simons variable $n=\mu/\pi$,
 \begin{equation}
 \label{Blocheq}
  \left( - {1 \over 2m_{\mu}}{\partial ^2 \over \partial {\mu}^2} +V(\mu) \right) \Psi(\mu) = i\frac{\partial\Psi(\mu)}{\partial t}
\end{equation} 
where the mass $m_{\mu}$ and the periodic potential $V(\mu)$ are given in Eq.(\ref{system}), with the barrier height $V_{max}=E_{sph} \simeq 9$ TeV. Here we need to clarify what the co-ordinate $\mu$ stands for with respect to the classical vacua and local minima that are present.
 
In the presence of left-handed fermions, the periodic potential direction we are interested in is different from the usual $|\theta\rangle$ vacuum direction. Here we first treat the fermions to be massless and so the local minima are more or less degenerate. The massive fermion case will be discussed afterwards. The direction for the Bloch waves is insensitive to whether the fermions are massive or not, though the masses of the fermions will lift the degeneracy of the periodic potential $V(\mu)$. For low-lying energies, this can make a significant difference. However, we are mostly interested in $E > E_{sph}$, and the fermion masses introduce only a small correction to the estimate of cross-sections, as we shall explain later.

 At the classical level, there exist $n_L=12$ ($i=1,2,...,n_L$) globally conserved $U(1)$ currents 
\begin{equation}
 \label{JLi}
J^{(i) \mu}={\bar \Psi}^{(i)}_L \gamma^{\mu}\Psi^{(i)}_L
\end{equation} 
corresponding to the conservation of the fermion numbers. However, this conservation is broken by the presence of anomaly \cite{Adler1969,Bell1969},
\begin{equation}
 \label{JLK}
\partial_{\mu} J^{(i) \mu} = \frac{g^2}{16 \pi^2} \Tr\left[F_{\mu \nu} \tilde{F}^{\mu \nu}\right] = \partial_{\mu} K^{\mu}
\end{equation} 
where $\tilde{F}^{\mu \nu}$ is the dual of  ${F}^{\mu \nu}$ and there exists a (non-gauge-invariant) current $K^{\mu}$. 
In the presence of instanton solutions in Euclidean space-time \cite{Belavin1975},  
\begin{equation}
 \label{chern} N =\frac{g^2}{16 \pi^2} \int d^4x \Tr\left[F_{\mu \nu} \tilde{F}^{\mu \nu} \right], 
\end{equation} 
where the topological index $N$ takes only integer values. An instanton with value $N$ leads to the tunneling process $\left|n \right\rangle \rightarrow \left|n+N \right\rangle$.

Combining the above results, we can construct a gauge variant conserved current ${\bar J}^{\mu}$ and the corresponding conserved charge ${\cal Q}$,
\begin{eqnarray}
&J_F^{\mu}=\frac{1}{n_L} \sum_{i=1}^{n_L}J^{(i) \mu}  \nonumber\\
&\partial_{\mu} {\bar J}^{\mu} = \partial_{\mu}(K^{\mu}-J_F^{\mu}) =0,   \nonumber\\
& {\cal Q} = \int d^3 x {\bar J}^0 =  Q_G -Q_F
\end{eqnarray} 
which is the winding number $Q_G$ of the gauge field minus the normalized baryon plus lepton (($B+L)/6$) number $Q_F$. 
Its change under gauge transformation $U$ with winding number one is 
\begin{equation}
 T_U{\cal Q}T_U^{-1} = {\cal Q} +1 \quad \Rightarrow \quad [T_U,{\cal Q}]=T_U,
\end{equation}
or $[T_U,Q_G]=T_U $. We also have the following commutation relations due to charge conservation and gauge invariance,
\begin{equation}
   [{\cal Q},H]=0, \quad [T_U,H]=0.
\end{equation}
So a state may be described by two values,
$|n\rangle =|n_G, n_F\rangle$, with $Q_G|n \rangle = n_G|n \rangle$ and $Q_F |n \rangle = n_F|n \rangle$, which has net baryon number (i.e., baryon minus anti-baryon number) $3n_F$ and net lepton number $3n_F$. 
Here 
\begin{equation}
\label{Qnumber}
 {\cal Q} |n\rangle =(n_G-n_F) |n_G, n_F\rangle = n|n\rangle
 \end{equation}
 so only the combined $n=n_G - n_F$ is conserved. Conservation of $n$ implies
\begin{equation}
  \langle n'|O(x_1)O(x_2)...|n \rangle \propto \delta(n-n') 
\end{equation}
 
Consider the classical vacuum states
$|n\rangle_0 = |n_G, n_F=0 \rangle$.  $T_U$ acts on them as a lowering operator. (Since they are vacua, they do not have baryon number, so they should also be eigenstates of $Q_G$, i.e. $Q_G|n\rangle_0=n|n\rangle_0$.) 
From this set of vacua one can construct the $\theta$-vacua, 
$$|\theta\rangle = \sum_n e^{in\theta}|n\rangle =  \sum e^{in_G\theta}|n_G,0\rangle ,$$
otherwise cluster decomposition is violated. 
Then one finds that all $|\theta \rangle$ have the same energy,
\begin{equation}
  H |\theta+\alpha\rangle = He^{i\alpha {\cal Q}}|\theta\rangle = e^{i\alpha {\cal Q}}E_\theta|\theta\rangle = E_\theta|\theta +\alpha\rangle.
\end{equation}
As the baryon number symmetry is exact at the Lagrangian level, performing such a rotation can rotate away the $\theta$ coupling in the effective Lagrangian. Therefore all $|\theta\rangle$ are equivalent. 
However, due to the super-selection rule of $\theta$-vacua in QFT, 
\begin{equation}
  \langle\theta'|O(x_1)O(x_2)...|\theta \rangle \propto \delta(\theta-\theta') 
\end{equation}
for all local operators $O(x)$.

For $n_F \ne 0$, the $|n\rangle =|n_G, n_F\rangle$ states 
are obviously not vacua, but these classical ground states are almost degenerate with the vacuum states for very soft massless fermions and for not too big $n_F$. 
The operator ${\cal Q}$ defines a set of ground states $|n\rangle$ by ${\cal Q}|n\rangle = n|n\rangle$ given in Eq.(\ref{Qnumber}).
The $\mu$ direction in the Schr\"odinger equation (\ref{Blocheq}) refers to any of the ${\cal Q}$ conserving direction. 
For example, a ($B+L$)-violating process with $\Delta n \ne 0$ refers to such a ${\cal Q}$ conserving transition: 
$$|n_G, n_F \rangle \rightarrow |n_G+\Delta n, n_F+\Delta n \rangle$$
For $n=0$, a Bloch state takes the form (for integer $\pi \mu \in {\bf Z}$),
$$|k \rangle= \sum_{\mu} e^{ik\mu}|\mu, \mu \rangle$$
where $\mu$ is the spatial QM coordinate in the Schr\"odinger equation (\ref{Blocheq}).
Clearly, $|k \rangle$ is very different from $|\theta \rangle$ in the presence of fermions.

 \section{Above the Sphaleron Height}\label{Above}
 
Following the Bloch Theorem, many interesting features are quite general, well captured in the solution to the one-dimensional time-independent Schr\"{o}dinger equation with mass ${\bar m}=m_{\mu} v^2$ and the periodic potential $V(q)$ (\ref{system}), where $q=\mu/v$. 
A wavefunction solution with dimensionless crystal momentum $k$ and energy $E(k)$ takes the form 
\begin{equation}
\label{Floquet}
\psi_k(\mu) =e^{ik\mu}u_k(\mu), \quad \quad u_k(\mu)=u_k(\mu+\pi)
\end{equation}
That is, up to a phase, $\psi(\mu)$ is periodic. 

Let us choose $k$ to lie in the first Brillouin zone (i.e., $-1 < k \le 1$), so $|K|=0, 2, 4, ...$ labels the bands starting from the Bloch wavefunction with the lowest energy. The spread of $E(k)$ gives the band width while the band gaps are given by the separation between bands. The solutions form the pass bands, which are separated by gaps without solution. The energies of the pass bands are essentially those of the bound states between any two barriers. 
As shown in Ref.\cite{tye:2015tva}, the lowest band width is about $10^{-180}$ TeV while the next band is about 70 GeV higher.  For low energy Bloch waves (small $|K|$), the magnitudes of $\psi_k(\mu)$ peak around $\mu=|K|\pi/2$ and are exponentially small around $\mu=(K+1)\pi/2$. In fact, $u_k(\mu)$ around the vacuum is essentially the bound state wavefunction there. With energies inside a pass band, the wave function spreads across the whole potential and transmission from one vacuum to another (at different integer $n$) is no longer tunneling suppressed.  However, the width of the bands at low energies, and so the mean velocity of motion, are exponentially small. Sitting at one minimum is in practice not different from being a bound state there. Averaging over a few bands and their gaps at low energies, we find that the probability to lie inside a band is exponentially suppressed. This is simply another way to see the tunneling suppression effect. 

\begin{figure}[h]
 \begin{center}
  \includegraphics[scale=0.4]{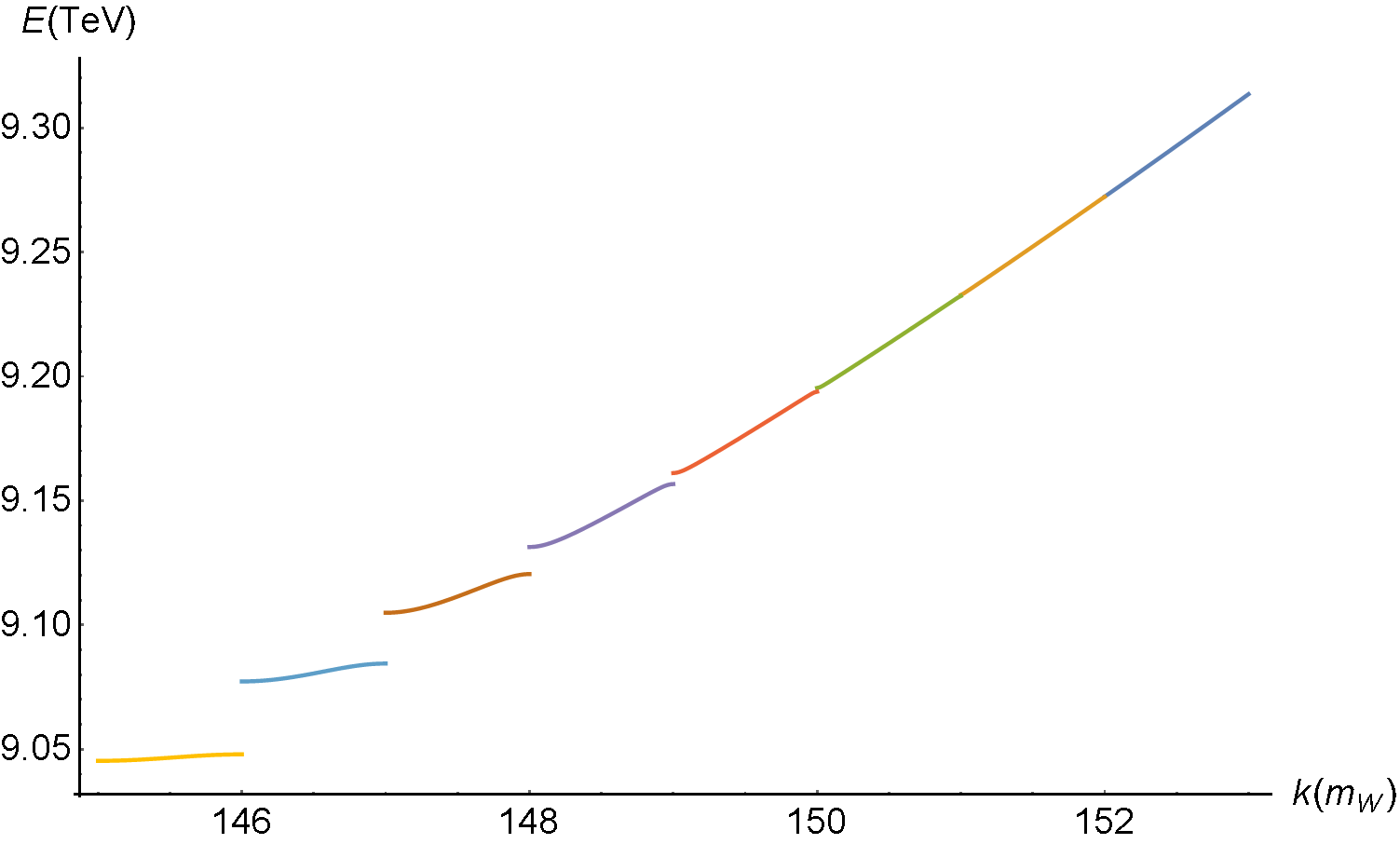}
  \caption{ A sketch of the extended Brillouin zone scheme for $E(k,K_1)$ versus $p=v(k+K_1)$. The height of the sphaleron potential is $E_{sph}=9.1$ TeV (before turning on the $U(1)$ hypercharge coupling). Note that the bands follow the parabolic curve ${\cal E}=p^2/2m$ for $E > E_{sph}=9.1$ TeV. Turning on the $U(1)$ hypercharge will lower the sphaleron mass from 9.1 TeV to 9.0 TeV.
  }
 \label{parabolic}
 \end{center}
\end{figure}

As energy increases, the bands become wider quickly while the band gaps decrease slowly.
As the energy goes above the sphaleron energy, $E> E_{sph}$, the solutions are very well approximated by plane waves, as shown in FIG \ref{parabolic}.  To avoid degeneracy, we should choose $k$ away from the edge eigenvalues: $k \ne 0, \pm 1$. 
The resulting wave function is a single mode of plane wave,
\begin{equation}
\label{e6}
\psi_k(K_1, \mu)  \simeq e^{i(k+K_1)\mu}
\end{equation} 
Once $E > E_{sph}$, the band gaps decrease rapidly while the band widths increase rapidly, so the band gaps are very narrow compared to the band widths, as shown in FIG \ref{parabolic}. As a result, a single plane wave mode is a very good approximation to the Bloch wave solution for $E > E_{sph}$.

That is, for energies above $E_{sph}$, the wavefunction (\ref{e6}) takes the time-dependent form
 \begin{equation}
 \label{plane}
\psi(q, t) \simeq \exp \left(ipq -i{\cal E}t \right)
\end{equation}
where the ``momentum" $p=v(k+K_1)$ is conjugate to the coordinate $q=\mu/v$, and the effective energy $\cal E$ depends on $E(k)$ and the average of the potential $V_0= 4.197$ TeV,
 \begin{equation}
 {\cal E} (p) = E(k,K_1) - V_0  = \frac{p^2}{2{\bar m}}= \frac{(k+K_1)^2}{2m_{\mu}}
 \end{equation}
  where ${\bar m}=m_{\mu}v^2$ has dimension of mass
and the mean velocity along the $\mu$ direction is given by
\begin{equation}
\label{meanv}
\left\langle {\hat v}\right\rangle = \left\langle \frac{\partial {\cal E}(p)}{\partial p}\right\rangle = \pm p/{\bar m}
\end{equation}
For $E(k,K_1) \gtrsim E_{sph}$, $|p| \simeq 13$ TeV. So we see that the ($B+L$)-violating processes are no longer tunneling suppressed when the energy is above $E_{sph}$. For $E(k,K_1) \gtrsim E_{sph}$, $|p| \gtrsim 40$ TeV. The time it takes to go from from $|n=0\rangle \rightarrow |n+1\rangle$ is of order of $10^{-25}$ sec.. 

In short, the Bloch wave is essentially a (single mode) plane wave, which is a solution of the periodic potential. So there is no tunneling suppression when $E > E_{sph}$, i.e., $F(E>E_{sph}) =0$, as illustrated by the blue curve in FIG \ref{transmission} for the one-dimensional QM case.

\section{Presence of Massive Fermions}\label{massfermion}

In the electroweak theory (with $v=246$ GeV), where the $\cal Q$ charge is conserved and the fermion zero modes suppress the $|n_G-1, n_F\rangle \rightarrow |n_G, n_F\rangle$ transition. Instead,  we have
the transition from $|n_G-1, n_F\rangle$ to
 $$ | n_G, n_F+1\rangle = \Pi_{i=1}^{12} \psi_{iL} |n_G, n_F\rangle \simeq BBBlll|n_G, n_F\rangle$$
Allowing the CKM mixing, the 3 baryons ($B$) plus 3 leptons ($l$) content can be $BBBlll = ppne \mu \nu_{\tau}$ which is about 3 GeV. Ignoring the CKM mixing, we have 3 $b$ quarks and 3 $s$ quarks, plus 3 $u$ (or $d$) quarks so $BBBlll$ contributes about 20 GeV. That is, the $|n_G, n_F+1 \rangle$ ground state is about $20$ GeV above the $|n_G-1, n_F \rangle$ ground state. Let us take the energy shift to be 
$$\delta E \simeq 20 \, {\rm GeV}$$
Now the sphaleron periodic potential $V_{sph}$ is modified to
\begin{equation}
\label{VC}
V(q) = V_{sph} (q) + C |q| = V_{sph} (q) + \frac{v \delta E}{\pi} |q|
\end{equation} 
where $q=n_G \pi/v=\mu/v$ and $C=v\delta E/\pi$ if we start at $n_G=q=0$.

Clearly, this is a big effect on the low-lying Brillouin bands. However, phenomenology is interesting only for energies above $E_{sph}=9$ TeV. So let us focus on that case. \\
\begin{figure}[h]
 \begin{center}
  \includegraphics[scale=0.6]{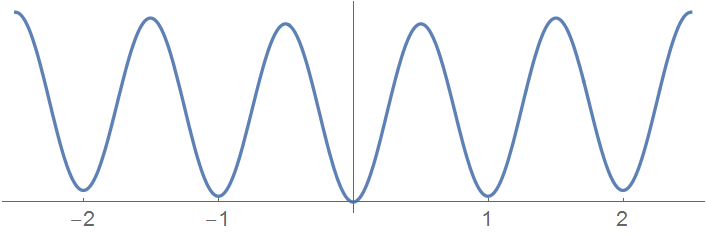}
 \end{center}
 \caption{A schematic plot of the potential (\ref{VC}) in the presence of fermions with masses. Assuming we are starting at $q=0$ with $Q_F=0$, fermion masses lift the potential minima away from zero for $q\ne 0$; that is, $C \ne 0$. }
\end{figure}
Turning on $C$ in the potential (\ref{VC}) means that $k$ and $E$ now take discrete values.
Due to the smallness of $C$ (i.e., $\delta E$), the spectrum is dense, so we can safely neglect this effect for $E > E_{sph}$. The presence of $C$ introduces only a slow variation to the potential.
The condition for the validity of the WKB approximation states that
$|dp/dq| \ll p^2$, or ${\bar m} C /|p| \ll  p^2$.
Near $q \sim 0$, we have 
\begin{equation}
\frac{{\bar m} C}{|p|^3} \simeq \frac{0.25 {\rm TeV}^3}{|p|^3} <  10^{-5} 
\end{equation}
so the change to the plane wavefunction (\ref{plane}) may be estimated via the WKB method, 
\begin{equation}
\psi(q)= \exp \left(ipq \right) \rightarrow \psi(q) \simeq \frac{1}{\sqrt{|p(q)|}}\exp \left( i\int^q p(q') dq' \right)
\end{equation}
where $p(q)^2 = 2 {\bar m} (E-V(q))>0$. This introduces a small effect around $q \sim 0$, when $E > V(q)$, but becomes big at $E \sim V(q)$. As an illustration, for $E \sim 10$ TeV, we have a wave function well approximated by     a plane wave until we get close to $|n_G| \sim 45$. So the motion from $n_G=0$ to $n_G=\pm 1$ is little changed by the introduction of fermion masses.

\section{The Quantum Mechanical Model} \label{twodim}

The discussion so far is for a one-dimensional QM problem.
Clearly, the QM model reduced from the electroweak theory has a number of relevant degrees of freedom, or dimensions. Besides the single ($B+L$)-violating dimension discussed above, we expect a number of ($B+L$)-conserving dimensions, or degrees of freedom (d.o.f.), as well.
In general, the impact of these d.o.f. is reserved for the pre-factor in the estimate of the tunneling rate. For energies above $E_{sph}$, when the tunneling suppression factor is absent in our estimate, the number of ($B+L$)-conserving d.o.f. becomes important in our estimate of the event rate. Since low energy properties in the Bloch wave analysis and the QM models studied in the literature mimicking the single sphaleron process are more or less in agreement with the numerical study \cite{Bezrukov2003}, we shall attempt to extract the number of ($B+L$)-conserving dimensions from such a study. Our main result is that, crudely speaking, there are of order $d \sim 4$ effective ($B+L$)-conserving dimensions. Here we shall first briefly review these QM models and adept them for our purpose to extract the information we are looking for, that is, the number of ($B+L$)-conserving d.o.f..

For a single potential barrier in the one-dimensional QM case, the problem is elementary. Starting with an incoming wave from the left of the localized potential barrier, we end up with a reflected wave on the left side and a transmitted wave on the right side.
If the energy of the incoming plane wave is below the barrier height, the transmission is tunneling suppressed.
For energies above the barrier height, the transmission probability is not suppressed and can even approach unity. 
This behavior is shown by the blue curve in FIG \ref{transmission}. 

 Recall the 2-dimensional QM model studied in Ref.\cite{Bonini:1999cn,Bonini:1999kj,Bezrukov:2003tg}. The goal there is to calculate the transmission probability for a single potential barrier. It is believed that this model captures some of the key qualitative properties of the ($B+L$)-violating process in the electroweak theory, including the ``few-to-many" suppression factor believed to be present. Here we briefly review the salient features of this model before we go to the periodic potential case. 
In this 2-dimensional QM model with a single potential barrier in one of the two directions, even if we start with an incoming wave with an energy along the barrier direction above the barrier height, coupling between the two directions will transfer energy to the ``barrier-free" direction so the energy in the barrier direction ends up below the barrier height and the transmission coefficient $T$ is still exponentially suppressed. 

We start with the following Lagrangian of two particles with the same mass (with $c= \hbar=1$), 
\begin{equation}
\label{Lagrangian1}
L =  \frac{m}{2}\dot{\mu}^2 + \frac{m}{2}\dot{y}^2 - \frac{1}{2}\omega^2 v^2\left(a\mu+ y\right)^2 - U(\mu)
\end{equation}
where $m=m_{\mu}$ (\ref{system}), $v=\frac{1}{\sqrt{1+a^2}}$ and $U(\mu)$ is a potential with a single barrier that mimics the sphaleron potential; 
 so the $\mu$ direction corresponds to the ($B+L$)-violating direction while $y$ direction corresponds to the ($B+L$)-conserving direction.  ($a=1$ corresponds to the 2-dim. case studied in Ref.\cite{Bonini:1999cn,Bonini:1999kj,Bezrukov:2003tg}.)
 For the single barrier case, we shall set the potential (\ref{system}) to zero everywhere except for $\pi > \mu \ge 0$; that is, 
 \begin{equation}
 U(\mu) =V(\mu)  \quad \quad  \pi > \mu \ge 0
 \end{equation}
with $U(\mu)=0$ otherwise. This problem is non-trivial because of their coupling in the simple harmonic oscillator, whose excitation mimics particle creation. WKB approximation is not straight forward in multi-dimensional problem. One has to identify the ``shortest path" of penetration through the barrier. In the case $\omega\to 0$, the shortest penetration is perpendicular to the barrier, which is along the $\mu$ direction. For energies $E_{\mu}$ below the barrier height, the transmission amplitude is approximately,
 \begin{align} \label{eqn:WKB}
 T &\sim \exp{\left(-\int d\mu \sqrt{2 m' (U-E_{\mu})} \right)} \nonumber \\
 &\rightarrow  \exp \left(-\frac{2 \pi}{\alpha_W} F_0(E_{\mu}(E))\right)
 \end{align}
in the limit of small coupling $g$ and low energy $E_{\mu}$, where $F_0$ is a finite constant of order unity. (In contrast to Ref.\cite{Bonini:1999cn,Bonini:1999kj,Bezrukov:2003tg}, here the mass $m'$ scales like $m' \propto g^{-3}$ while $U \propto 1/g$). So we see that the standard coupling dependence emerges. As a result, the transmission coefficient $T$ is exponentially suppressed for low incoming energies and small coupling. The dependence $E_{\mu}$ on total incoming energy $E$ is determined below.
 
One starts with an incoming plane wave with energy $E$ while sitting at the ground state of the harmonic oscillator mimicking small initial particle numbers. Both the reflected and the transmitted waves will have energies transferred to the $y$ direction, exciting the harmonic oscillator.
As a result, even for $E>E_{sph}$, because of the energy being drained to the $y$ direction, the energy available to go over the barrier may be less than the barrier height, so tunneling is necessary, thus resulting in a tunneling suppression. For high enough energies, there is no tunneling suppression, so $F_0(E, \omega) \rightarrow 0$. The key result is shown in FIG \ref{transmission}, in which we reproduce the function $F_0(E, \omega)$ as a function of $E$ as obtained in Ref.\cite{Bonini:1999cn,Bonini:1999kj,Bezrukov:2003tg}. 

As $\omega \rightarrow \infty$, excitation of the harmonic oscillator needs a huge amount of energy, so the harmonic oscillator stays in the ground state and no energy is transferred to the $y$ direction. In this case, $F_0(E, \omega) \rightarrow 0$ as $E \rightarrow  E_{sph}$ (shown by the blue curve in FIG \ref{transmission}). As $\omega \rightarrow 0$, excitations of the harmonic oscillator needs little energy, so it is easy to transfer energy to the $y$ direction. In this $a=1$ case, $F_0(E, \omega) \rightarrow 0$ as $E \rightarrow  2E_{sph}$. For finite $\omega$, $F_0(E, \omega)$ will follow a curve in between the above 2 curves.

To study the energy dependence of $F_0(\omega, E)$, it is convenient to go to a different choice of orthonormal basis: $x_1=v(\mu+ay)$ and $x_2=v(a\mu-y)$, in which the coupling between the coordinates in the harmonic oscillator is transferred to the ``sphaleron" potential,
\begin{align}
\label{lagrangian3}
L &=  \frac{ m}{2}\dot{x_1}^2 + \frac{ m}{2}\dot{x_2}^2 - \frac{1}{2}\omega^2 x_2^2 - U(v(ax_1-x_2)).
\end{align}
We shall see, $d \sim a^2$ mimics the effective number of  number of ($B+L$)-conserving directions. The assumption that very few oscillation mode are excited means initial kinetic energy is in the $x_1$-direction. As we see that the biggest effect of the ($B+L$)-conserving direction is in the $\omega \rightarrow 0$ limit, let us explore this a little further. (Note that this is different from the case where the harmonic oscillator term is absent from the beginning.) In this case we may neglect the harmonic oscillator term. To get some intuition about the transferring of energy in the other direction, we calculate the initial energy of the $\mu$ particle,
\begin{align}
\label{lagrange mu new}
E_{\mu} &= \frac{m}{2} \dot{\mu}^2 
= \frac{m}{2 (1 + a^2)}  \left(\dot{x_1}^2 +  a^2 \dot{x_2}^2 + 2 a \dot{x_1} \dot{x_2}\right) \nonumber\\
&= \frac{E}{(1 + a^2)}  \left( 1 + a^2 \frac{E_{x_2}}{E} + 2 a \sqrt{\left(\frac{E_{x_2}}{E}\right)}  \right) \simeq \frac{E}{(1 + a^2)} \nonumber
\end{align}
where in the last step, we have assumed that the initial energy is mostly in the direction of $x_1$. 

With the dependence of $E_{\mu}(E)$, we can use Eq.(\ref{eqn:WKB}) to get the behavior of the transmission amplitude as shown in Fig. \ref{transmission}. Here we essentially reproduce their result as expected : the blue curve is for $a=1$ and $\omega \rightarrow \infty$ where $F_0 =0$ at $E=E_{sph}$. For $a=1$ and $\omega \rightarrow 0$,  $F_0 =0$ at $E=2E_{sph}$. For finite $\omega$, we expect curves in between these 2 limiting cases.
The orange curve in FIG  \ref{transmission} is for$a^2=2$ and $\omega \rightarrow 0$ where $F_0 =0$ at $E=3E_{sph}$. In general, for $\omega \rightarrow 0$, $F_0 \rightarrow 0$ as $E \rightarrow (1+a^2)E_{sph}$. Crudely speaking $d=a^2$ measures the number of ($B+L$)-conserving directions. 

As $\omega \rightarrow 0$, the coupling between the ($B+L$)-violating and ($B+L$)-conserving directions weakens while exciting the harmonic oscillating modes becomes easier; this results in an easier energy transfer to the baryon conserving direction and a higher incoming energy $E$ is needed to avoid tunneling suppression. That is, to avoid tunneling suppression, for $\omega \rightarrow 0$,
\begin{equation}
E > (1+a^2) E_{sph}
\end{equation}



\section{A Crude Estimate of $\kappa (E_{qq} > E_{sph})$} \label{kappa1}

For the periodic sphaleron potential, $F(E)$ follows the blue curve in FIG {\ref{transmission}. Even without the tunneling suppression factor for $E_{qq} > E_{sph}$, there can still be some suppression coming from the pre-factor due to the ($B+L$)-conserving d.o.f.. Here we  like to guessestimate the number of ($B+L$)-conserving directions that accompany the ($B+L$)-violating direction $q$ and estimate the effective pre-factor, or $\kappa$.
For low energy, we have
\begin{equation}
\label{lowek}
\kappa \simeq \frac{\rm band width}{{\rm band gap}+{\rm band width}} \sim e^{-4\pi F(0)/\alpha_W}
 \end{equation}
which is comparable to the standard estimate as $F(E=0) \sim 1$.
For $E_{qq} > E_{sph}$, $F(E)=0$, while the pre-factor is present due to phase space considerations; that is, not all $E_{qq}$ goes in the ($B+L$)-violating direction.
Let us assume that the system can be modeled by a $(d+1)$-dimensional QM problem,
\begin{equation}
L = \frac{p_{q}^2}{2m} + \sum_{i=1}^{d} \frac{p_i^2}{2m_i} - V(q,x_i),
\end{equation}
where $p_{q}$ is the $q$ direction for baryon number violation and $p_i$ correspond to the baryon number conserving $x_i$ directions; the latter directions are not expected to encounter any potential barrier. For energies above $E_{sph}$ in the $q$ direction, we have plane waves in all directions. We further assume that all masses are the same and the initial state has no preferred momentum direction (the direction of momentum is equally distributed in all directions). The initial state with energy $E$ stays at the bottom of the potential. The chance of having enough energy to overcome the barrier is given by the portion of phase space on the surface of $d$-sphere (${\vec{p}}^2 = 2E_{qq}$) with $|p_q| > \sqrt{2E_{sph}}$. This is just twice the solid angle in $d$-dim with half cone angle $\alpha=\cos ^{-1} \sqrt{{E_{sph}}/{E_{qq}}}$, 
\begin{align}
  &\frac{2}{S^d}\int dS^{d-1} \int_0^{\alpha}\sin^{d-1}\phi d\phi   \nonumber \\
  = &~ 1- \frac{2 \Gamma((d+1)/2)}{ \sqrt{\pi}\Gamma(d/2)}
 \sqrt{\frac{E_{sph}}{E_{qq}}}  ~_2F_1\left(\frac{1}{2}, \frac{2-d}{2}, \frac{3}{2}, \frac{E_{sph}}{E_{qq}} \right) \nonumber
\end{align}
For $E_{sph}/E_{qq} \lesssim 1$, we have 
\begin{equation}
\label{phasespace}
 \kappa (E_{qq}) \sim \frac{1}{d}\left (2-2\sqrt{\frac{{E_{sph}}}{E_{qq}}}\right)^{d/2}
 \end{equation} 
for some constant $d$. Crudely speaking, $d$ may be interpreted as the number of the ($B+L$)-conserving directions. Comparing this with Ref.\cite{Bezrukov2003} for regions where they agree, we have (see FIG \ref{transmission}), 
\begin{equation} 
d \simeq  a^2 \simeq 4
\end{equation}
as a reasonable guesstimate of the effective number of ($B+L$)-conserving dimensions. A somewhat different value of $d$ is also perfectly acceptable to us. It is important that $d$ is not a large number. 
That is, the probability of sending most energy to the ($B+L$)-violating direction is suppressed by phase space consideration. We may choose other forms, but it is reasonable to assume that the suppression is power-like, not exponential-like. For $d \simeq 4$, $E_{qq} \gtrsim 9.5$ TeV where $E_{qq}> E_q >9.0$ TeV, $\kappa \sim 10^{-3}$, as quoted in Eq.(\ref{estimatek}). Admittedly, this estimated value of $\kappa$ may have large uncertainties even if the tunneling suppression factor is absent. The key point is that this is a phase space suppression factor, which should be power-like, not a tunneling suppression factor, which would have been be exponential. In short, this value of $\kappa$ (quoted in Ref.\cite{tye:2015tva}) is much larger than the exponentially small value given in Eq.(\ref{oldk}).

\section{Discussion}\label{discussion}

The discrepancy in the two very different conclusions on the ($B+L$)-violating cross-section for $E>E_{sph}$ (\ref{oldk},\ref{estimatek}) may be seen in either the QFT viewpoint or the reduced QM approach. Both involves the question whether tunneling involving multiple sphaleron barriers is multiply suppressed or not, and furthermore, whether there are situations where tunneling through multiple identical barriers may be more efficient than tunneling through a single barrier. Earlier analyses implicitly assume that tunneling over multiple sphalerons will be multiply suppressed, so they can be safely ignored. We argue otherwise.

In the electroweak theory, the coupling of a W-boson to a sphaleron involves an inverse power of the coupling $g$. 
FIG \ref{WtoSph} gives a simple picture where resonance effect can, under the appropriate circumstances, enhance the double sphaleron event rate with respect to the single sphaleron event rate. However, a direct QFT calculation of the rates of such processes is non-trivial.

If we formulate a QFT problem in the functional Schr\"odinger formulation, it can be reduced to a QM system with some number of relevant degrees of freedom. In a simple tunneling problem in QFT, a single QM degree of freedom dominates the exponential factor, while the remaining QM d.o.f. contribute to the pre-factor of the tunneling rate. This is the picture we adopt. The phase space estimate for the pre-factor carried out in the estimate of $\kappa (E > E_{sph})$ in Sec. \ref{kappa1} is based on this reasonable assumption.

Assuming that the ($B+L$)-conserving QM d.o.f. can only contribute to the pre-factor in the cross-section, the red dash curve in FIG {\ref{transmission} may be interpreted in the following way. The number $d$ of ($B+L$)-conserving d.o.f. is actually energy dependent; the number $d(E)$ increases rapidly as we increase the energy  to $E \gg E_{sph}$. The resulting pre-factor is so big that it can actually impact on the exponent to reach Eq.(\ref{2tomany}).  

In the periodic potential case in the QM system, the number $d$ of ($B+L$)-conserving d.o.f. is fixed, independent of energy. When the Bloch wave along the sphaleron direction is a simple plane wave for $E > E_{sph}$, it is no different from the plane waves in the ($B+L$)-conserving directions, so the phase space distribution is probabilistic, as suggested in the above analysis. That is, there is no exponential tunneling suppression above the sphaleron energy, and the small number of  ($B+L$)-conserving d.o.f. cannot modify the exponent factor; they contribute to the pre-factor only as a phase space factor, which is power-like.

The above discussions are restricted to the QM analogue for the sphaleron physics in the electroweak theory.
Notice that in both cases, namely the single barrier case and the periodic potential case, we start with a plane wave in the $\mu$-direction (to be identified with the baryon number violating direction). In the single barrier case \cite{Bonini:1999cn,Bonini:1999kj,Bezrukov:2003tg}, the incoming plane wave is an initial condition and the resulting solution to the system is a combination of the incident wave plus a reflected wave on one side of the potential barrier and a transmitted wave on the other side. Because the scattered waves pick up energy in the ($B+L$)-conserving direction, the energy available to go over the barrier is decreased and so tunneling suppression is still present when the energy $E_{qq}$ is not high enough, even if we start with $E_{qq}>E_{sph}$. 
That is, the ($B+L$)-conserving d.o.f. is still coupled to the ($B+L$)-violating d.o.f so the former can modify the exponent factor in the tunneling rate. 
In the periodic potential case, when $E_{qq} > E_{sph}$, to a good approximation the incoming plane wave is a Bloch wave solution of the full potential already. So there is no further scattering and no tunneling suppression, though some kinematic (power-like) suppression is expected due to the contribution by ($B+L$)-conserving d.o.f. to the pre-factor. 

It has been argued \cite{Khoze1991,Khoze1991a} that, as the incoming energy increases, the ($B+L$)-violating event rate may grow rapidly in the single sphaleron case, due to the increasing number of gauge bosons being produced; so such events may be observed for incoming proton-proton energy $E \sim 100$ TeV; but the event rate is still too small to be observable at around $E \sim 40$ TeV. The dynamics of such an enhancement of the ($B+L$)-violating cross-section is very different from our Bloch wave argument, which suggests that such events should already be observable at $E \sim 30$ TeV. Of course, their ``many boson" effect may play a secondary role in our event rate estimate.

\section{Summary} \label{summary}

The electroweak theory is very well-defined and being a weak coupling theory, it is surprising that the very interesting ($B+L$)-violating processes are still poorly understood. Recently we proposed that the existing estimate of the cross-sections of such processes in the prevalent picture are off (too small) by many orders of magnitude. We agree with the existing estimate for a single sphaleron, but we argue that multi-sphaleron processes are very important, due to the resonance phenomenon. For a periodic potential with a discrete translational symmetry (in the $\mu$ direction), this phenomenon is built in in the Bloch wave solutions.

In our picture, the cross section is tunneling suppressed at low energies, $\sigma \sim \exp [-4\pi/\alpha_W]$ as given in Eq.(\ref{lowek}), which is in agreement with the prevalent picture. As we increase the energy to above the sphaleron energy, the tunneling suppression factor disappears, so the ($B+L$)-violating cross-sections are lifted to an observable level in the near future. Still, there are at least two other factors come into play : \\
\{1\} The presence of the ($B+L$)-conserving directions which will take up some of the incoming 2-particle energies.
In our QM picture, they can contribute only to the pre-factor of the scattering event rate, resulting in a power-like phase space suppression factor (\ref{estimatek}). This is different from the earlier QM analysis which allow the ($B+L$)-conserving directions to impact on the exponential factor in the event rate, leading to the bound (\ref{oldk}).\\
\{2\} Since the parton distribution function and the phase space factor in $\kappa$ severely limit these cross-sections for the LHC 14 TeV run, a run at a higher proton-proton energy, say $E  \sim 25$ TeV, will substantially ameliorate this parton distribution function suppression and the phase space suppression in $\kappa$ so it should provide a definitive check on whether such cross-sections are tunneling suppressed or not.
Note also that $E_{sph}$ will be lower for a flatter Higgs potential than the standard quartic form; hence such a flattening will increase the chances of discovery. Clearly experimental measurements of the Higgs couplings will be important to determine the form of the Higgs potential.
 
In the earlier analyses, which assume multi-sphaleron processes are totally negligible compared to the single sphaleron process, the cross-section also increases as we increase the energy. However, as energy increases even to $E \gg E_{sph}$, the cross-section asymptotes to $\sigma \sim \exp [-2\pi/\alpha_W]$ due to the ``few-to-many" suppression. In our picture, this suppression, though present, is erased by the resonance effect mentioned in FIG \ref{WtoSph}. This phenomenon is present only in the multi-sphaleron picture in Lorentzian time, which is captured by our Bloch wave analysis.  

Although our estimate of a much bigger rate is subject to some (albeit reasonable) assumptions, the discrepancy strongly suggests that a renewed study of the ($B+L$)-violating processes is clearly warranted. It will be a challenge to re-examine this multi-sphaleron issue in the electroweak theory entirely within the field theory framework. Also, although the ``few-to-many" factor should be present in the production of a soliton, it remains to be clarified whether it should be present (and to what level) when we are tunneling through a sphaleron barrier.

\vspace{3mm}


 {\bf Acknowledgments} - 
 We are grateful to Andy Cohen for lengthy discussions which are of great value to this work. 
We also thank John Ellis, Razieh Emami and Kazuki Sakurai for useful discussions. The work is supported by the CRF Grants of the Government of the Hong Kong SAR under HKUST4/CRF/13G and the GRF 16305414 issued by the Research Grants Council (RGC) of Hong Kong. The research of SW is supported in part by the Croucher Foundation. 

\vspace{4mm}

\appendix

\section{The Resonant Tunneling Phenomenon}

Resonant tunneling is a well understood phenomenon in quantum mechanics \cite{merzbacher1998quantum}. However, it is less appreciated in QFT. Here we review a simple old argument why one expects it to be present in QFT as well. Consider the tunneling from state $A$ to state $C$ via an intermediate state $B$, as shown in FIG \ref{twoB}. To simplify the argument, let the tunneling rate $\Gamma_{A \rightarrow B}  = \Gamma_{B \rightarrow C} \simeq e^{-S}$, where $S \gg 1$. If one rotates to Euclidean time and evaluate the $A \rightarrow C$ tunneling rate,
one finds  $\Gamma_{A \rightarrow C} \sim e^{-2S}$. 

\begin{figure}[h]
 \begin{center}
  \includegraphics[scale=0.4]{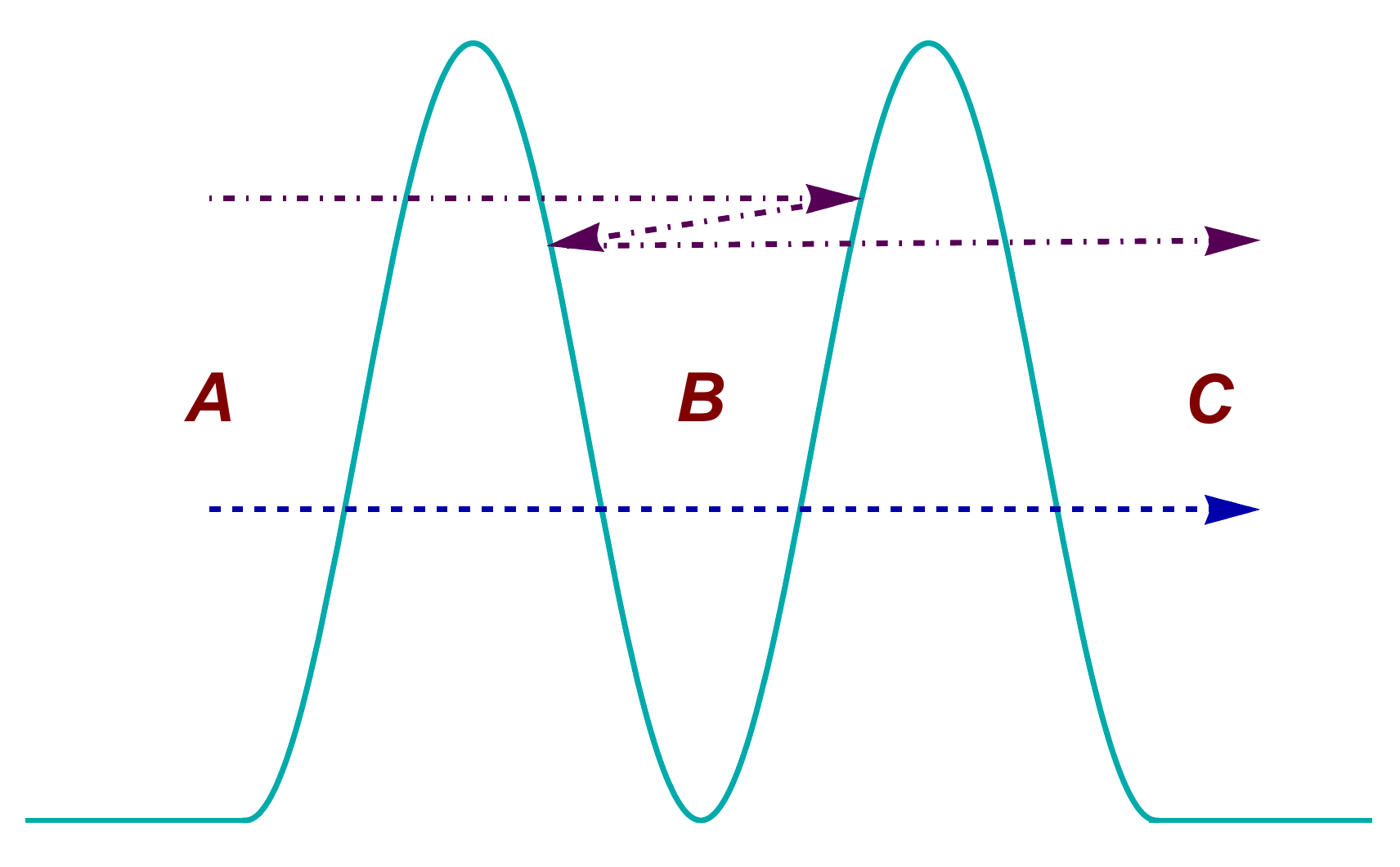}
  \caption{The 3 regions, $A$, $B$ and $C$, separated by two identical potential barriers. The dashed lines are 2 of the possible paths in the path integral going from $A$ to $C$. If, for the right energy, the extra round trip in region $B$ attains a $2\pi$ phase for the top path, there are infinitely many paths with multiple round trips in region $B$ contributing to a coherent sum. This is the resonant tunneling phenomenon, which can overcome the exponential tunneling suppression to yield a transmission coefficient $T \rightarrow 1$ in QM. }
  \label{twoB}
 \end{center}
\end{figure}

Now consider the time it takes to tunnel from $A \rightarrow C$, namely
$t(A  \rightarrow  C)=1/\Gamma_{A \rightarrow C}$, which is equal to the time it takes to tunnel from $A \rightarrow B$ ($= 1/\Gamma_{A \rightarrow B}$) plus the time
it takes to tunnel from $B  \rightarrow  C$ ($= 1/ \Gamma_{B \rightarrow C}$),
$$t(A  \rightarrow  C) = t(A  \rightarrow  B) + t(B  \rightarrow  C) $$
Since this total tunneling time gives the inverse of the tunneling rate  $\Gamma_{A\rightarrow C}$, we have
\begin{equation} \label{A1}
\frac{1}{\Gamma_{A\rightarrow C}}= \frac{1}{\Gamma_{A \rightarrow B}} + \frac{1}{\Gamma_{B \rightarrow C}}
\end{equation}
That is, 
\begin{equation}
\label{ABC}
\Gamma_{A \rightarrow C} = \frac{\Gamma_{A \rightarrow B}\Gamma_{B \rightarrow C}}{\Gamma_{A \rightarrow B}+\Gamma_{B \rightarrow C}} \simeq  e^{-S}
\end{equation}
That is, the $A \rightarrow C$ tunneling is singly suppressed, not doubly suppressed. This result should be true in both QM and QFT. It is easy to extend the argument to see that the tunneling through multiple barriers is again only singly suppressed, not multiply suppressed. 

How to reach such a tunneling rate ? Clearly the $e^{-2S}$ (for $S \gg 1$) contribution is far from sufficient.  In QM, we understand that this averaged rate (\ref{ABC}) is due to the resonant tunneling effect; that is, when we hit a resonance (or bound state) in $B$ (see FIG \ref{twoB}). 

In one-dimensional QM with a single potential barrier, a plane wave hitting the barrier will have reflected and transmitted waves with reflection probability $|R|^2$ and transmission probability $|T|^2$ so that $|R|^2+|T|^2=1$. Here, $|T|$ is tunneling suppressed for energies below the barrier height and unsuppressed for energies above the barrier height. This picture agrees more or less with the case when the potential is periodic.  In the two identical barriers case as shown in FIG \ref{twoB}, the transmission coefficient $T \rightarrow 1$ (i.e., saturates the unitarity bound) when the energy hits the bound state (resonance) value. At other energies, $T$ is doubly suppressed. 
In terms of the path integral, this means that whenever a round trip in B has a phase of an integer multiple of $2 \pi$, a coherent sum of (such multiple round trips) paths enhances the tunneling rate. This is the resonant tunneling phenomenon. Since any resonance in $B$ has a decay width of order $e^{-S}$, the transmission rate averages to $\Gamma \sim e^{-S}$. Clearly the $e^{-2S}$ contribution is totally negligible here. 

The above argument (\ref{ABC}) should apply in QFT as well. To get such a fast average (i.e., the $e^{-S}$) rate in QFT, resonant tunneling should be present as well. The FIG \ref{WtoSph} picture strongly suggests that it is not hard to reach such a situation in the multi-sphaleron tunneling processes. However, a QFT calculation may be challenging. An easy way to capture this effect in QFT is to first reduce the system to a quantum mechanical system, as is carried out here. For a periodic potential, Bloch waves capture the essence of the resonant tunneling phenomenon.

Since the resonant tunneling is due to the coherent sum of paths for bound states (resonances) in the intermediate state, the usual (naive) rotation to Euclidean time in the evaluation of the tunneling rate will miss this important phenomenon.

\bibliographystyle{utphys}
\bibliography{References}

\end{document}